\pdfoutput=1
\documentclass[11pt,a4paper]{article}
\usepackage[hyperref]{emnlp2020}
\usepackage{times}
\usepackage{latexsym}
\usepackage{soul}
\usepackage{url}
\usepackage{graphicx}
\usepackage{amsmath}
\usepackage{booktabs}
\usepackage{latexsym}
\usepackage{microtype}
\usepackage{subfigure}
\usepackage{graphicx}
\usepackage{amssymb}
\usepackage[boxed,ruled,commentsnumbered]{algorithm2e}
\usepackage{multirow}
\usepackage{makecell}
\usepackage{verbatim}
\usepackage{changepage}
\usepackage{xcolor}

\usepackage{microtype}

\aclfinalcopy

\def\aclpaperid{2323}

\title{PP-Rec: News Recommendation with Personalized User Interest \\and Time-aware News Popularity}

\author{Tao Qi$^1$, Fangzhao Wu$^2$, Chuhan Wu$^1$ and Yongfeng Huang$^1$\\
  $^1$Department of Electronic Engineering \& BNRist, Tsinghua University, Beijing 100084, China  \\
  $^2$Microsoft Research Asia, Beijing 100080, China\\
  {\tt \{taoqi.qt, wufangzhao, wuchuhan15\}@gmail.com}\\
  {\tt yfhuang@mail.tsinghua.edu.cn}\\}

\date{}

\begin{document}
\maketitle

\begin{abstract}

Personalized news recommendation methods are widely used in online news services.
These methods usually recommend news based on the matching between news content and user interest inferred from historical behaviors.
However, these methods usually have difficulties in making accurate recommendations to cold-start users, and tend to recommend similar news with those users have read.
In general, popular news usually contain important information and can attract users with different interests.
Besides, they are usually diverse in content and topic.
Thus, in this paper we propose to incorporate news popularity information to alleviate the cold-start and diversity problems for personalized news recommendation.
In our method, the ranking score for recommending a candidate news to a target user is the combination of a personalized matching score and a news popularity score.
The former is used to capture the personalized user interest in news.
The latter is used to measure time-aware popularity of candidate news, which is predicted based on news content, recency, and real-time CTR using a unified framework.
Besides, we propose a popularity-aware user encoder to eliminate the popularity bias in user behaviors for accurate interest modeling.
Experiments on two real-world datasets show our method can effectively improve the accuracy and diversity for news recommendation.


\end{abstract}

\section{Introduction}

Personalized news recommendation is a useful technique to help users alleviate information overload when visiting online news platforms~\cite{wu2020mind,wuuser,wu2020fairness,ge2020graph}.
Existing personalized news recommendation methods usually recommend news to a target user based on the matching between the content of candidate news and user interest inferred from previous behaviors~\cite{danzhu2019,wu2019pd}.
For example, \citet{wu2019NRMS} proposed to model news content from news title based on multi-head self-attention.
In addition, they modeled user interest from the previously clicked news articles with multi-head self-attention to capture the relatedness between different behaviors.
\citet{an2019neural} proposed to use CNN network to learn news embeddings from news titles and categories, and model both long-term and short-term user interests from news click behaviors.
However, these personalized news recommendation methods usually have difficulties in making accurate recommendations to cold-start users, since the behaviors of these users are very sparse and it is difficult to model their interest~\cite{trevisiol2014cold}.
Besides, these methods tend to recommend similar news with those users have read~\cite{nguyen2014exploring}, which may hurt user experience and is not beneficial for them to receive new information.


\begin{figure}
    \centering
    \resizebox{0.48\textwidth}{!}{\includegraphics{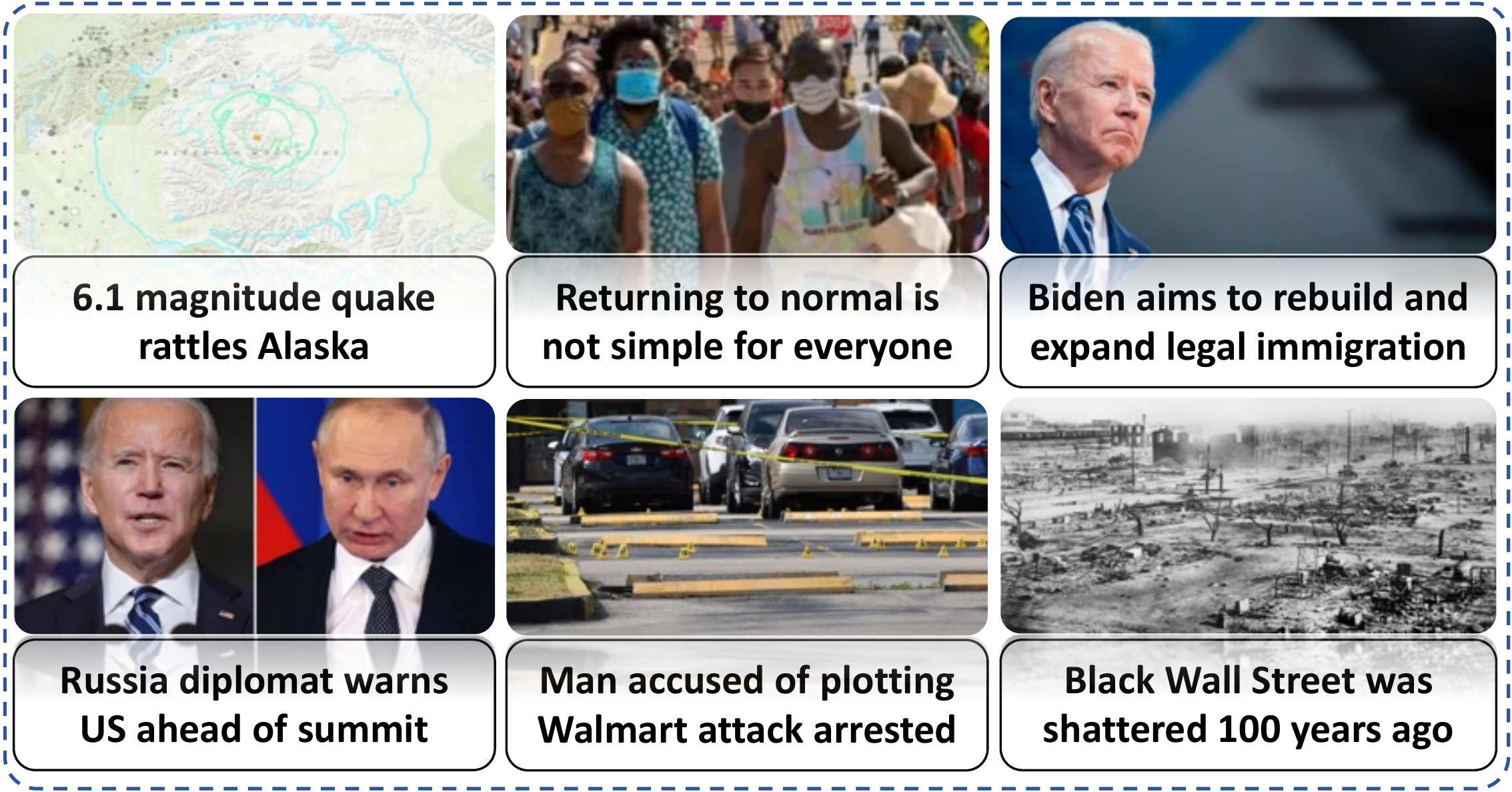}}
    \caption{Several example popular news.}

    \label{fig.motivation}
    \vspace{-0.1in}
\end{figure}

The motivation for this work is that popular news usually convey important information such as catastrophes, epidemics, presidential election and so on, as shown in Fig.~\ref{fig.motivation}.
These popular news can attract many users to read and discuss even if they have different personal interest~\cite{yang2016effects}.
In addition, popular news are diverse in content and can cover many different topics~\cite{houidi2019news}.
Thus, incorporating popular news has the potential to alleviate the cold-start and diversity problems in personalized news recommendation.

In this paper, we propose a new method named \textit{PP-Rec} for news recommendation\footnote{ https://github.com/JulySinceAndrew/PP-Rec}, which can consider not only personalized user interest in news but also the popularity of candidate news.
In our method, the ranking score of recommending a candidate news to a target user is the combination of a personalized matching score and a news popularity score.
The personalized matching score is used to measure personal user interest in the content of candidate news.
The news popularity score is used to measure the time-aware popularity of candidate news.
Since news popularity is influenced by many different factors such as content and freshness, we propose a unified model to predict time-aware news popularity based on news content, recency, and near real-time click-through rate (CTR). 
These two scores are combined via a personalized aggregator for news ranking, which can capture the personalized preferences of different users in popular news.
Moreover, we propose a knowledge-aware news encoder to generate news content embeddings from both news texts and entities.
Besides, since news popularity can effect users' click behaviors~\cite{zheng2010clicks} and lead to bias in behavior based user interest modeling, we propose a popularity-aware user encoder which can consider the popularity bias in user behaviors and learn more accurate user interest representation.
Extensive experiments on two real-world datasets show \textit{PP-Rec} can effectively improve the performance of news recommendation in terms of both accuracy and diversity.

\section{Related Work}
\subsection{Personalized News Recommendation}

Personalized news recommendation are widely used in online news platforms~\cite{liu2010personalized,bansal2015content,wu2020mind,wu2020ptum,wu2019neurald}.
Existing personalized news recommendation methods usually rank candidate news for a target user based on the matching between news content and user interest~\cite{wang2018dkn,qi2020privacy,wu2020sentirec,wutanr}.
For example, \citet{okura2017embedding} learned news embeddings from news bodies via an auto-encoder and modeled user interests from the clicked news via a GRU network.
The matching between news and user is formulated as the dot product of their embeddings.
\citet{wu2019NRMS} used multi-head self-attention networks to generate news content embeddings from news titles and generate user interest embeddings from clicked news.
They also used the dot product of user and news embeddings as personalized matching scores for news ranking.
These personalized news recommendation methods usually model user interests from previous news click behaviors.
However, it is difficult for these methods to make accurate recommendation to cold-start users whose behaviors are very sparse~\cite{trevisiol2014cold}.
These users are very common in online news platforms, making the cold-start problem become a critical issue in real systems~\cite{sedhain2014social}.
Although some methods were proposed to alleviate the cold-start problem in personalized recommendation~\cite{sedhain2014social,trevisiol2014cold}, they usually utilized side information~\cite{son2016dealing} such as social network~\cite{lin2014personalized} to enhance user interest modeling.
However, the side information used in these methods may be unavailable in news recommendation.
In addition, these personalized methods tend to recommend similar news with those users have already read, which makes it difficult for users to receive new news information and may hurt their news reading experience~\cite{nguyen2014exploring,wu2019pd}.
Different from these methods, in \textit{PP-Rec} we consider not only users' personal interest in news but also the popularity of candidate news, which can alleviate both cold-start and diversity problems to some extent.

\subsection{Popularity-based News Recommendation}

\begin{figure*}
    \centering
    \resizebox{0.75\textwidth}{!}{
    \includegraphics[clip]{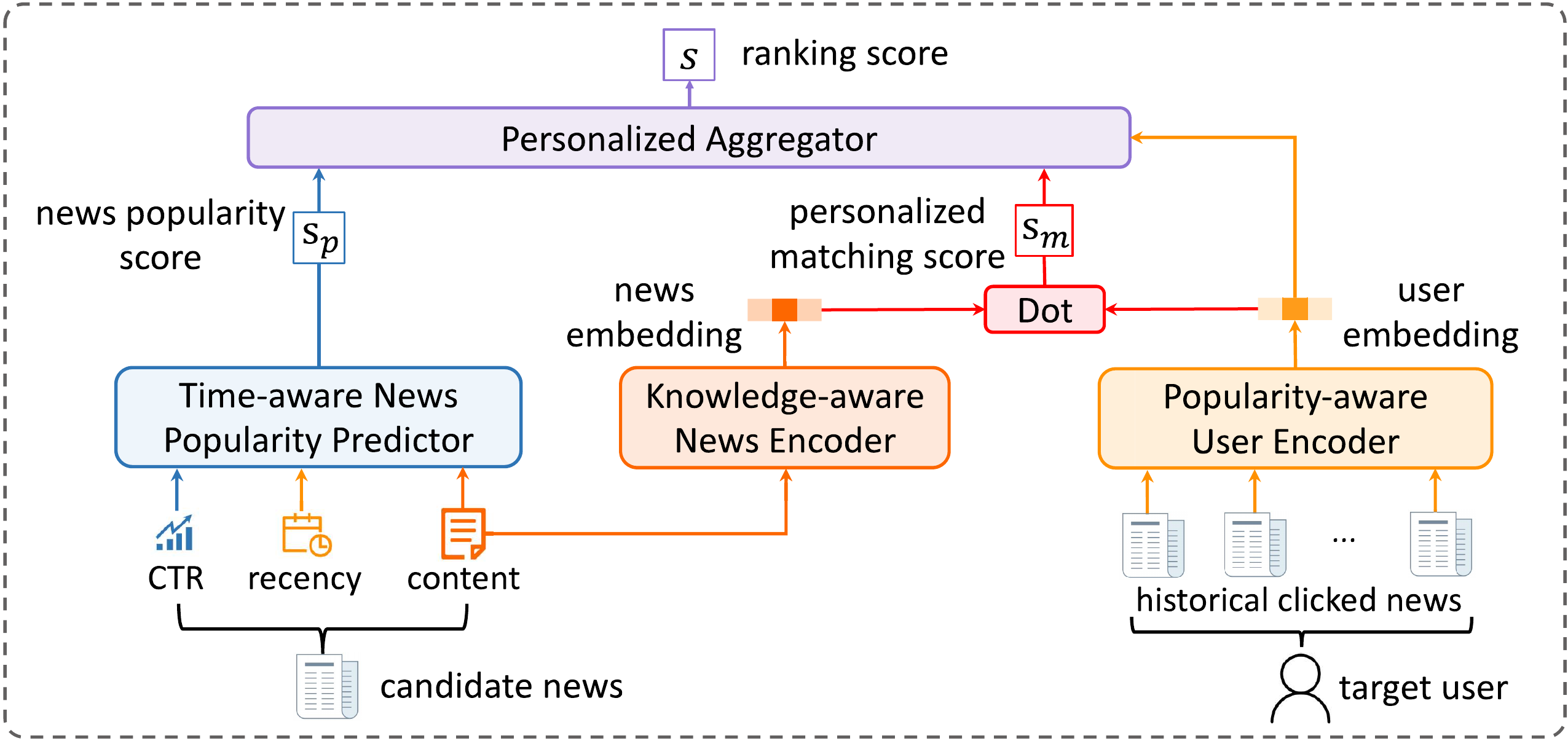}
    }
    \caption{The overall framework of \textit{PP-Rec}.}
    \label{fig.framework}
\end{figure*}

Our work is also related to popularity-based news recommendation methods.
Different from personalized news recommendation methods which rank candidate news based on users' personal interests, popularity-based news recommendation methods rank candidate news based on their popularity~\cite{phelan2009using,tatar2014popularity,lerman2010using,szabo2010predicting,jonnalagedda2016incorporating}.
A core problem in popularity-based news recommendation methods is how to estimate the popularity of candidate news accurately.
Most existing methods estimated news popularity based on the statistics of users' interactions with news on online news platforms, such as the number of views and comments~\cite{yang2016effects,tatar2014popularity,lee2010approach}.
For example, \citet{yang2016effects} proposed to use the frequency of views to measure news popularity.
\citet{tatar2014popularity} proposed to predict news popularity based on the number of comments of news via a linear model.
\citet{li2011scene} proposed to use the number of clicks on news to model their popularity and further adjust the ranking of news with same topics based on their popularity.
However, different news usually have significant differences in impression opportunities, and these view and comment numbers are biased by impression times.
Different from these methods, we use CTR to model news popularity, which can eliminate the impression bias.
Besides CTR, we also incorporate the content and recency information of candidate news to predict the popularity of candidate news in a more comprehensive and time-aware manner.


\section{Methodology}

\begin{figure}
    \centering
    \resizebox{0.49\textwidth}{!}{
    \includegraphics[clip]{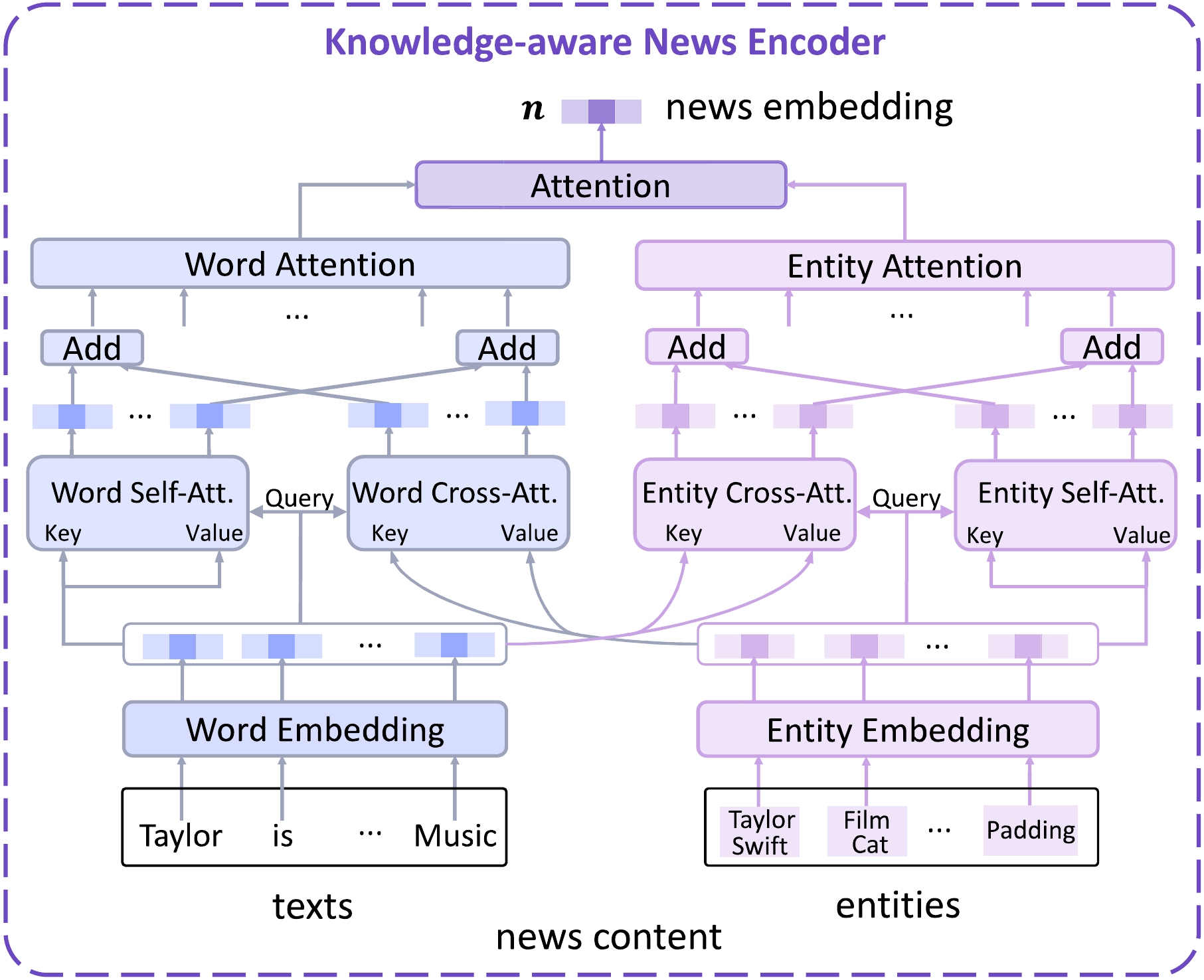}
    }
    \caption{Knowledge-aware news encoder.}
    \label{fig.news_encoder}
    

\end{figure}

In this section, we introduce \textit{PP-Rec} for news recommendation which can consider both the personal interest of users and the popularity of candidate news.
First, we introduce the overall framework of \textit{PP-Rec}, as shown in Fig.~\ref{fig.framework}.
Then we introduce the details of each module in \textit{PP-Rec}, which are shown in Figs.~\ref{fig.news_encoder},~\ref{fig.predictor} and~\ref{fig.model.user_encoder}.

    


\subsection{Framework of PP-Rec}

In \textit{PP-Rec}, the ranking score of recommending a candidate news to a target user is the combination of a personalized matching score $s_m$ and a news popularity score $s_p$.
The personalized matching score is used to measure the user's personal interest in the content of candidate news, and is predicted based on the relevance between news content embedding and user interest embedding.
The news content embedding is generated by a \textit{knowledge-aware news encoder} from both news texts and entities.
The user interest embedding is generated by a \textit{popularity-aware user encoder} from the content of clicked news as well as their popularity.
The news popularity score is used to measure the time-aware popularity of candidate news, which is predicted by a \textit{time-aware news popularity predictor} based on news content, recency, and near real-time CTR.

\subsection{Knowledge-aware News Encoder}
\label{sec:news_encoder}

First, we introduce the \textit{knowledge-aware news encoder}, which is shown in Fig.~\ref{fig.news_encoder}.
It learns news representation from both text and entities in news title.
Given a news title, we obtain the word embeddings based on word embedding dictionary pre-trained on large-scale corpus to incorporate initial word-level semantic information.
We also convert entities into embeddings based on pre-trained entity embeddings to incorporate knowledge information in knowledge graphs to our model.

There usually exists relatedness among entities in the same news.
For example, the entity ``MAC'' that appears with the entity ``Lancome'' may indicate cosmetics while it usually indicates computers when appears with the entity ``Apple''.
Thus, we utilize an entity multi-head self-attention network~\cite{vaswani2017attention} (MHSA) to learn entity representations by capturing their relatedness.
Besides, textual contexts are also informative for learning accurate entity representations.
For example, the entity ``MAC'' usually indicates computers if its textual contexts are ``Why do MAC need an ARM CPU?'' and indicates cosmetics if its textual contexts are ``MAC cosmetics expands AR try-on''.
Thus, we propose an entity multi-head cross-attention network (MHCA) to learn entity representations from the textual contexts.
Then we formulate the unified representation of each entity as the summation of its representations learned by the MHSA and MHCA networks.
Similarly, we use a word MHSA network to learn word representations by capturing the relatedness among words and a word MHCA network to capture the relatedness between words and entities.
Then we build the unified word representation by adding its representations generated by the word MHSA and the word MHCA networks.

Since different entities usually contribute differently to news representation, we use an entity attention network to learn entity-based news representation $\textbf{e}$ from entity representations.
Similarly, we use a word attention network to learn word-based news representation $\textbf{w}$ from word representations.
Finally, we learn the unified news representation $\textbf{n}$ with a weighted combination of $\textbf{e}$ and $\textbf{w}$ via an attention network.


\begin{figure}
    \centering
    \resizebox{0.415\textwidth}{!}{
    \includegraphics[clip]{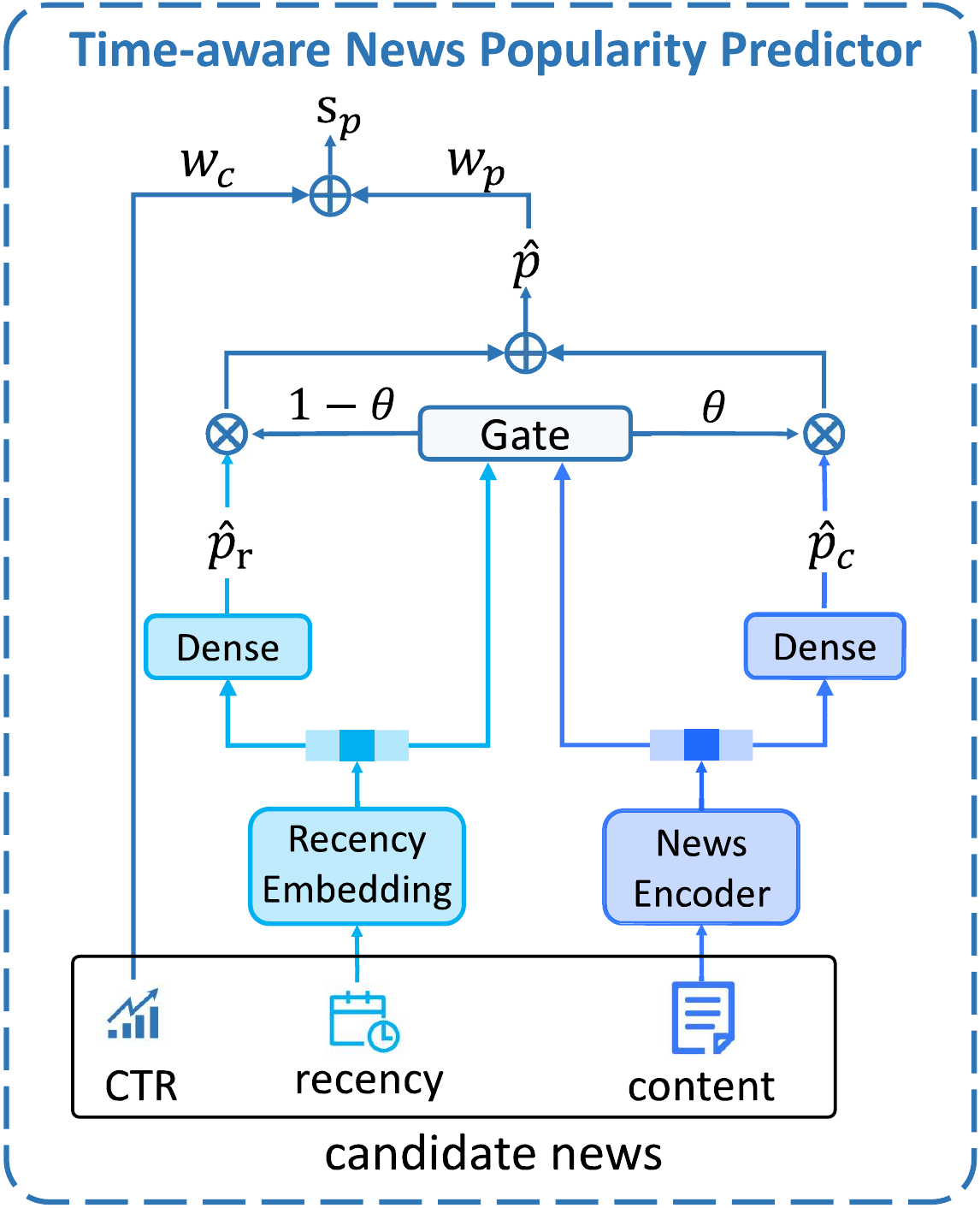}
    }
    \caption{Time-aware news popularity predictor.}
    \label{fig.predictor}
    

\end{figure}

\subsection{Time-aware News Popularity Predictor}

Next, we introduce the \textit{time-aware news popularity predictor}, as shown in Fig.~\ref{fig.predictor}.
It is used to predict time-aware news popularity based on news content, recency, and near real-time CTR information.
Since popular news usually have a higher click probability than unpopular news, CTR can provide good clue for popular news~\cite{jiang2016research}.
Thus, we incorporate CTR into news popularity prediction.
Besides, popularity of a news article usually dynamically changes.
Popular news may become less popular as they get out-of-date over time.
Thus, we use user interactions in recent $t$ hours to calculate near real-time CTR (denoted as $c_t$) for news popularity prediction.
However, the accurate computation of CTR needs to accumulate sufficient user interactions, which is challenging for those newly published news.

Fortunately, news content is very informative for predicting news popularity.
For example, news on breaking events such as earthquakes are usually popular since they contain important information for many of us.
Thus, besides near real-time CTR, we incorporate news content into news popularity prediction.
We apply a dense network to the news content embedding $\textbf{n}$ to predict the content-based news popularity $\hat{p}_c$.
Since news content is time-independent and cannot capture the dynamic change of news popularity, we incorporate news recency information, which is defined as the duration between the publish time and the prediction time.
It can measure the freshness of news articles, which is useful for improving content-based popularity prediction.
We quantify the news recency $r$ in hours and use a recency embedding layer to convert the quantified news recency into an embedding vector $\textbf{r}$.
Then we apply a dense network to $\textbf{r}$ to predict the recency-aware content-based news popularity $\hat{p}_r$.
Besides, since different news content usually have different lifecycles, we propose to model time-aware content-based news popularity $\hat{p}$ from $\hat{p}_c$ and $\hat{p}_r$ using a content-specific aggregator: 
\begin{equation}
 \hat{p} = \theta\cdot \hat{p}_c+(1-\theta)\cdot \hat{p}_r, \ \theta = \sigma( \textbf{W}^p\cdot [\textbf{n},\textbf{r}] + \textbf{b}^p ),
\end{equation}
where $\theta\in(0,1)$ means the content-specific gate, $\sigma(\cdot)$ means the sigmoid activation, $[\cdot,\cdot]$ means the concatenation operation, $\textbf{W}^p$ and $\textbf{b}^p$ are the trainable parameters.
Finally, the final time-aware news popularity $s_{p}$ is formulated as a weighted summation of the content-based popularity $\hat{p}$ and the CTR-based popularity $c_t$, i.e., $s_{p} = w_{c}\cdot c_t+w_{p}\cdot \hat{p}$,
where $w_{c}$ and $w_{p}$ are the trainable parameters.

\begin{figure}
    \centering
    \resizebox{0.415\textwidth}{!}{
    \includegraphics[clip]{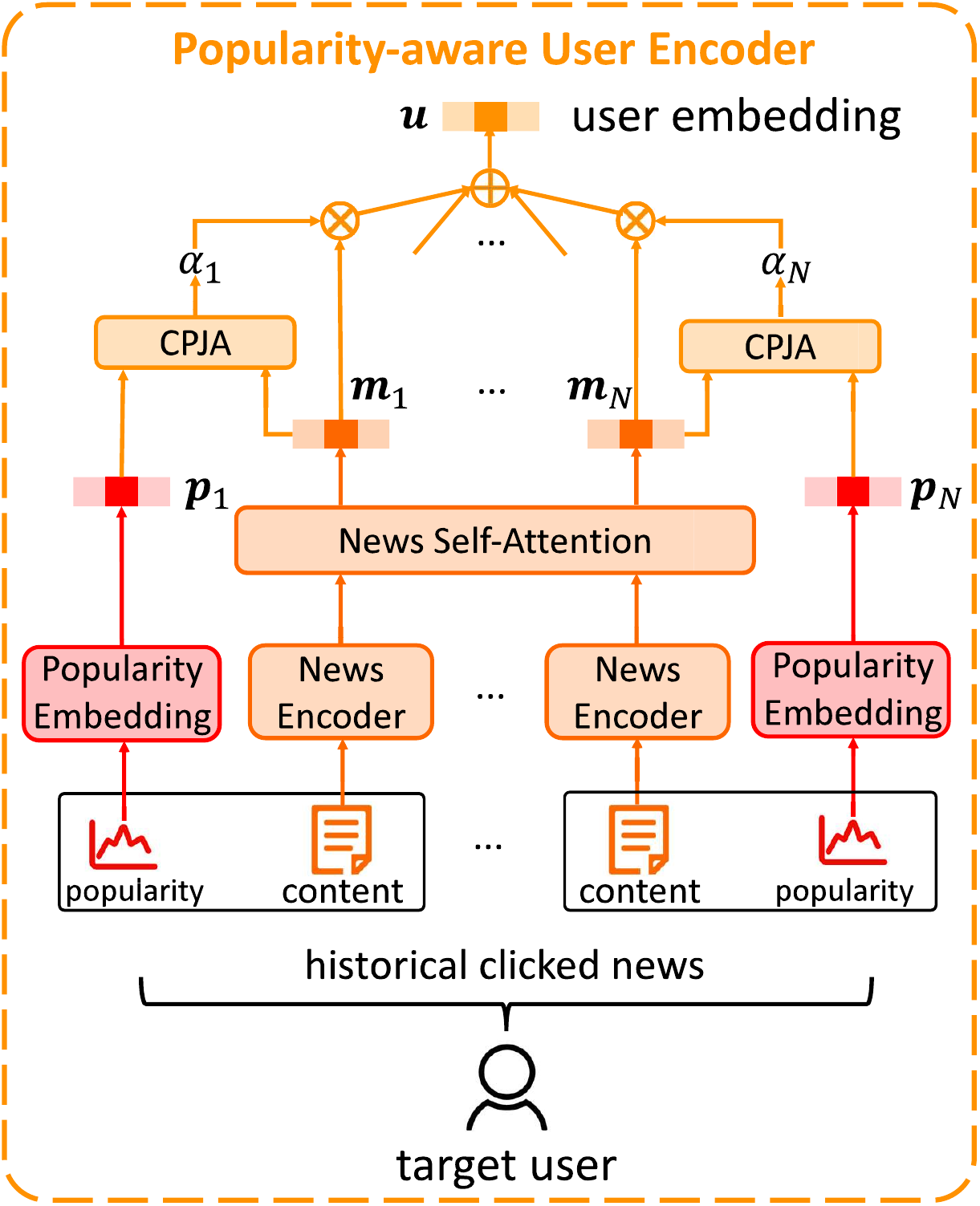}
    }
    \caption{Popularity-aware user encoder.}
    \label{fig.model.user_encoder}
    

\end{figure}

\subsection{Popularity-aware User Encoder}

Next, we introduce the \textit{popularity-aware user encoder} in \textit{PP-Rec} for user interest modeling, which is shown in Fig.~\ref{fig.model.user_encoder}.
In general, news popularity can influence users' click behaviors, and causes bias in behavior based user interest modeling~\cite{zheng2010clicks}.
Eliminating the popularity bias in user behaviors can help more user interest from user behaviors more accurately.
For example, a user may click the news ``Justin Timberlake unveils the song'' because he likes the songs of ``Justin Timberlake'', while he may click the news ``House of Representatives impeaches President Trump'' because it is popular and contains breaking information.
Among these two behaviors, the former is more informative for modeling the user interest.
Thus, we design a \textit{popularity-aware user encoder} to learn user interest representation from both content and popularity of clicked news.
It contains three components, which we will introduce in details.

First, motivated by~\citet{wu2019NRMS}, we apply a news multi-head self-attention network to the representations of clicked news to capture their relatedness and learn contextual news representation.
Second, we uniformly quantify the popularity of the $i$-th clicked news predicted by the \textit{time-aware news popularity predictor}\footnote{We remove news recency and content here to avoid non-differentiable quantization operation.} and convert it into an embedding vector $\textbf{p}_i$ via popularity embedding.
Third, besides news popularity, news content is also useful for selecting informative news to model user interest~\cite{wu2019ijcai}.
Thus, we propose a content-popularity joint attention network (CPJA) to alleviate popularity bias and select important clicked news for user interest modeling, which is formulated as:
\begin{equation}
    \alpha_i = \frac{\exp( \textbf{q}^T\cdot \tanh(\textbf{W}^u\cdot [\textbf{m}_i,\textbf{p}_i]) )}{\sum_{j=1}^N \exp( \textbf{q}^T\cdot \tanh(\textbf{W}^u\cdot [\textbf{m}_j,\textbf{p}_j]) )   } ,
\end{equation}
where $\alpha_i$ and $\textbf{m}_i$ denote the attention weight and the contextual news representation of the $i$-th clicked news respectively. $\textbf{q}$ and $\textbf{W}^u$ are the trainable parameters.
The final user interest embedding $\textbf{u}$ is formulated as a weighed summation of the contextual news representations: $ \textbf{u} = \sum_{i=1}^N \alpha_i\cdot \textbf{m}_i$.

\subsection{News Ranking and Model Training}

In this section, we introduce how we rank the candidate news and train the model in detail.
The ranking score of a candidate news for a target user is based on the combination of a personalized matching score $s_m$ and a news popularity score $s_p$.
The former is computed based on the relevance between user embedding $\mathbf{u}$ and news embedding $\mathbf{n}$.
Following~\citet{okura2017embedding}, we adopt dot product to compute the relevance.
The latter is predicted by the \textit{time-aware news popularity predictor}.
In addition, the relative importance of the personalized matching score and the news popularity score is usually different for different users.
For example, the news popularity score is more important than the personalized matching score for cold-start users since the latter is derived from scarce behaviors and is usually inaccurate.
Thus, we propose a personalized aggregator to combine the personalized matching score and news popularity score:
\begin{equation}
     s = (1- \eta)\cdot s_{m} + \eta\cdot s_{p},
\end{equation}
where $s$ denotes the ranking score, and the gate $\eta$ is computed based on the user representation $\textbf{u}$ via a dense network with sigmoid activation.

We use the BPR pairwise loss~\cite{rendle2009bpr} for model training.
In addition, we adopt the negative sampling technique to select a negative sample for each positive sample from the same impression.
The loss function is formulated as:
\begin{equation}
    \mathcal{L} = -\frac{1}{|\mathcal{D}|}\sum_{i=1}^{|\mathcal{D}|} \log(\sigma(s^p_i-s^n_i)),
\end{equation}
where $s^p_i$ and $s^n_i$ denote the ranking scores of the $i$-th positive and negative sample respectively, and $\mathcal{D}$ denotes the training dataset.


\section{Experiment}

\subsection{Dataset and Experimental Settings}


To our best knowledge, there is no off-the-shelf news recommendation dataset with news popularity information.
Thus, we built two datasets by ourselves.
The first one is collected from the user logs in the Microsoft News website from October 19 to November 15, 2019, and is denoted as \textit{MSN}.
We use the user logs in the last week for evaluation and others for model training and validation.
The second dataset is collected from a commercial news feeds in Microsoft from January 23 to April 23, 2020, and is denoted as \textit{Feeds}.
We use the logs in the last three weeks for evaluation and the rest for model training and validation.
For both datasets, we randomly sample 500k impressions for model training, 100k impressions for validation, and 500k impressions for evaluation, respectively.
The detailed statistics are listed in Table~\ref{table.stat}.
Following previous works~\cite{wu2019ijcai,an2019neural}, we use AUC, MRR, nDCG@5, and nDCG@10 to evaluate recommendation performance.

\begin{table}[h]
\centering
\resizebox{0.48\textwidth}{!}{
\begin{tabular}{ccccc}
\hline
                          & \# News & \# Users & \# Impressions & \#  Clicks \\ \hline
\textit{MSN}              & 161,013   & 490,522   & 1,100,000  & 1,675,084  \\
\textit{Feeds}            & 4,117,562   & 98,866  &  1,100,000 & 2,384,976  \\
\hline
\end{tabular}
}
\caption{Statistics of the datasets.}
\label{table.stat}
\end{table}

In our experiments, word embeddings are 300-dimensional and initialized by the Glove embeddings~\cite{pennington2014glove}.
The entity embeddings are 100-dimensional vectors pre-trained on knowledge tuples extracted from WikiData via TransE~\cite{bordes2013translating}.
We use clicked and unclicked impressions in the recent one hour to compute the near real-time CTR.
The recency and popularity embeddings are set to 100 dimensions and initialized randomly.
All multi-head attention networks are set to have 20 attention heads and the output dimension of each head is 20.
All gate networks are implemented by a two-layer dense network with 100-dimensional hidden vectors.
Dropout approach~\cite{srivastava2014dropout} is applied to \textit{PP-Rec} to migrate overfitting.
The dropout probability is set to 0.2.
Adam~\cite{kingma2014adam} is used for model training with $10^{-4}$ learning rate.
Hyper-parameters of \textit{PP-Rec} and baselines are tuned based on the validation set.

\subsection{Performance Evaluation}

We compare \textit{PP-Rec} with two groups of baselines.
The first group is popularity-based news recommendation methods, including:
(1) \textit{ViewNum}~\cite{yang2016effects}: using the number of news view to measure news popularity;
(2) \textit{RecentPop}~\cite{ji2020re}: using the number of news view in recent time to measure news popularity;
(3) \textit{SCENE}~\cite{li2011scene}: using view frequency to measure news popularity and adjusting the ranking of news with same topics based on their popularity;
(4) \textit{CTR}~\cite{ji2020re}: using news CTR to measure news popularity.
The second group is personalized news recommendation methods, containing:
(1) \textit{EBNR}~\cite{okura2017embedding}: utilizing an auto-encoder to learn news representations and a GRU network to learn user representations;
(2) \textit{DKN}~\cite{wang2018dkn}: utilizing a knowledge-aware CNN network to learn news representations from news titles and entities;
(3) \textit{NAML}~\cite{wu2019ijcai}: utilizing attention network to learn news representations from news title, body and category;
(4) \textit{NPA}~\cite{wu2019npa}: utilizing personalized attention networks to learn news and user representations;
(5) \textit{NRMS}~\cite{wu2019NRMS}: utilizing multi-head self-attention networks to learn both news and user representations;
(6) \textit{LSTUR}~\cite{an2019neural}: modeling users' short-term interests via the GRU network and long-term interests via the user ID;
(7) \textit{KRED}~\cite{liu2020kred}: learning news representation from titles and entities via a knowledge graph attention network.

\begin{table*}[t]
\centering
\resizebox{0.98\textwidth}{!}{

\begin{tabular}{|c|cccc|cccc|}
\hline
\multirow{2}{*}{Methods} & \multicolumn{4}{c|}{\textit{MSN}}                                                            & \multicolumn{4}{c|}{\textit{Feeds}}                                                               \\ \cline{2-9} 
                         & \textbf{AUC}            & \textbf{MRR}            & \textbf{nDCG@5}         & \textbf{nDCG@10}       & \textbf{AUC}            & \textbf{MRR}            & \textbf{nDCG@5}         & \textbf{nDCG@10}        \\ \hline
ViewNum                  & 54.12$\pm$0.00          & 24.95$\pm$0.00          & 26.07$\pm$0.00          & 31.56$\pm$0.00         & 58.99$\pm$0.00          & 23.71$\pm$0.00          & 26.83$\pm$0.00          & 32.38$\pm$0.00          \\
RecentPop                & 55.67$\pm$0.00          & 28.72$\pm$0.00          & 30.45$\pm$0.00          & 36.62$\pm$0.00         & 56.27$\pm$0.00          & 24.93$\pm$0.00          & 28.37$\pm$0.00          & 33.89$\pm$0.00          \\
SCENE                  & 57.89$\pm$0.02          & 27.41$\pm$0.01         & 28.81$\pm$0.02          & 34.36$\pm$0.03        & 60.82$\pm$0.03          & 27.29$\pm$0.03          & 31.25$\pm$0.02         & 36.56$\pm$0.03          \\
CTR                      & 65.72$\pm$0.00          & 30.50$\pm$0.00          & 32.79$\pm$0.00          & 38.68$\pm$0.00         & 66.40$\pm$0.00          & 30.29$\pm$0.00          & 35.53$\pm$0.00          & 40.72$\pm$0.00          \\\hline
EBNR                      & 63.90$\pm$0.20          & 30.13$\pm$0.12          & 32.25$\pm$0.14          & 38.05$\pm$0.14         & 64.88$\pm$0.04          & 28.91$\pm$0.03          & 33.29$\pm$0.03          & 38.87$\pm$0.02          \\
DKN                      & 64.16$\pm$0.19          & 30.63$\pm$0.10          & 32.98$\pm$0.12          & 38.66$\pm$0.11         & 66.30$\pm$0.11          & 30.25$\pm$0.06          & 35.01$\pm$0.07          & 40.55$\pm$0.06          \\
NAML                     & 66.06$\pm$0.17          & 32.10$\pm$0.10          & 34.73$\pm$0.11          & 40.43$\pm$0.11         & 67.50$\pm$0.09          & 31.07$\pm$0.08          & 36.08$\pm$0.10          & 41.61$\pm$0.10          \\
NPA                      & 65.83$\pm$0.20          & 31.70$\pm$0.09          & 34.24$\pm$0.10          & 39.96$\pm$0.10         & 67.25$\pm$0.10          & 30.80$\pm$0.05          & 35.72$\pm$0.07          & 41.25$\pm$0.07          \\
NRMS                     & 66.34$\pm$0.16          & 32.00$\pm$0.08          & 34.68$\pm$0.09          & 40.39$\pm$0.09         & 68.10$\pm$0.05          & 31.47$\pm$0.03          & 36.61$\pm$0.03          & 42.12$\pm$0.03          \\
LSTUR                    & 66.69$\pm$0.16          & 32.12$\pm$0.05          & 34.76$\pm$0.05          & 40.51$\pm$0.04         & 67.43$\pm$0.16          & 30.95$\pm$0.11          & 35.92$\pm$0.16          & 41.45$\pm$0.14          \\ 
KRED                      & 66.54$\pm$0.17          & 31.97$\pm$0.14          & 34.65$\pm$0.14          & 40.38$\pm$0.14         & 67.67$\pm$0.18          & 31.16$\pm$0.13          & 36.19$\pm$0.16          & 41.72$\pm$0.16          \\
\hline
PP-Rec                 & \textbf{71.05}$\pm$0.09 & \textbf{39.34}$\pm$0.08 & \textbf{44.01}$\pm$0.13 & \textbf{50.46}$\pm$0.20 & \textbf{72.11}$\pm$0.21 & \textbf{32.42}$\pm$0.12 & \textbf{38.13}$\pm$0.08 & \textbf{43.50}$\pm$0.13 \\ \hline
\end{tabular}

}

\caption{News recommendation results of different methods. We perform t-test and the results show that \textit{PP-Rec} significantly outperforms other baseline methods at significance level $p<0.001$. }
\label{Table.PE}

\end{table*}

We repeat each experiment 5 times and show average performance and standard deviation in Table~\ref{Table.PE}, from which we have the following observations.
First, among the popularity-based news recommendation methods, the \textit{CTR} method outperforms the \textit{ViewNum} method.
This is because the number of news views is influenced by impression bias while CTR can eliminate the impression bias and better measure news popularity.
Second, \textit{PP-Rec} outperforms all popularity-based methods.
This is because these methods usually recommend popular news to different users.
However, different users might prefer different news according to their personalized interests, some of which are not popular and cannot be recommended by these popularity-based methods.
In contrast, \textit{PP-Rec} considers both popularity and personalization in news recommendation.
Third, \textit{PP-Rec} outperforms all personalized methods.
This is because personalized methods usually recommend news based on the matching between news and user interest inferred from users' clicked news, and they ignore the popularity of each news.
However, popular news usually contain important and eye-catching information and can attract the attention of many users with different interests.
Different from these personalized methods, \textit{PP-Rec} incorporates news popularity into personalized news recommendation, which can recommend popular news to users and improve the performance of news recommendation.

\subsection{Performance on Cold-Start Users}



\begin{figure}
    \flushleft
    \resizebox{0.46\textwidth}{!}{\includegraphics{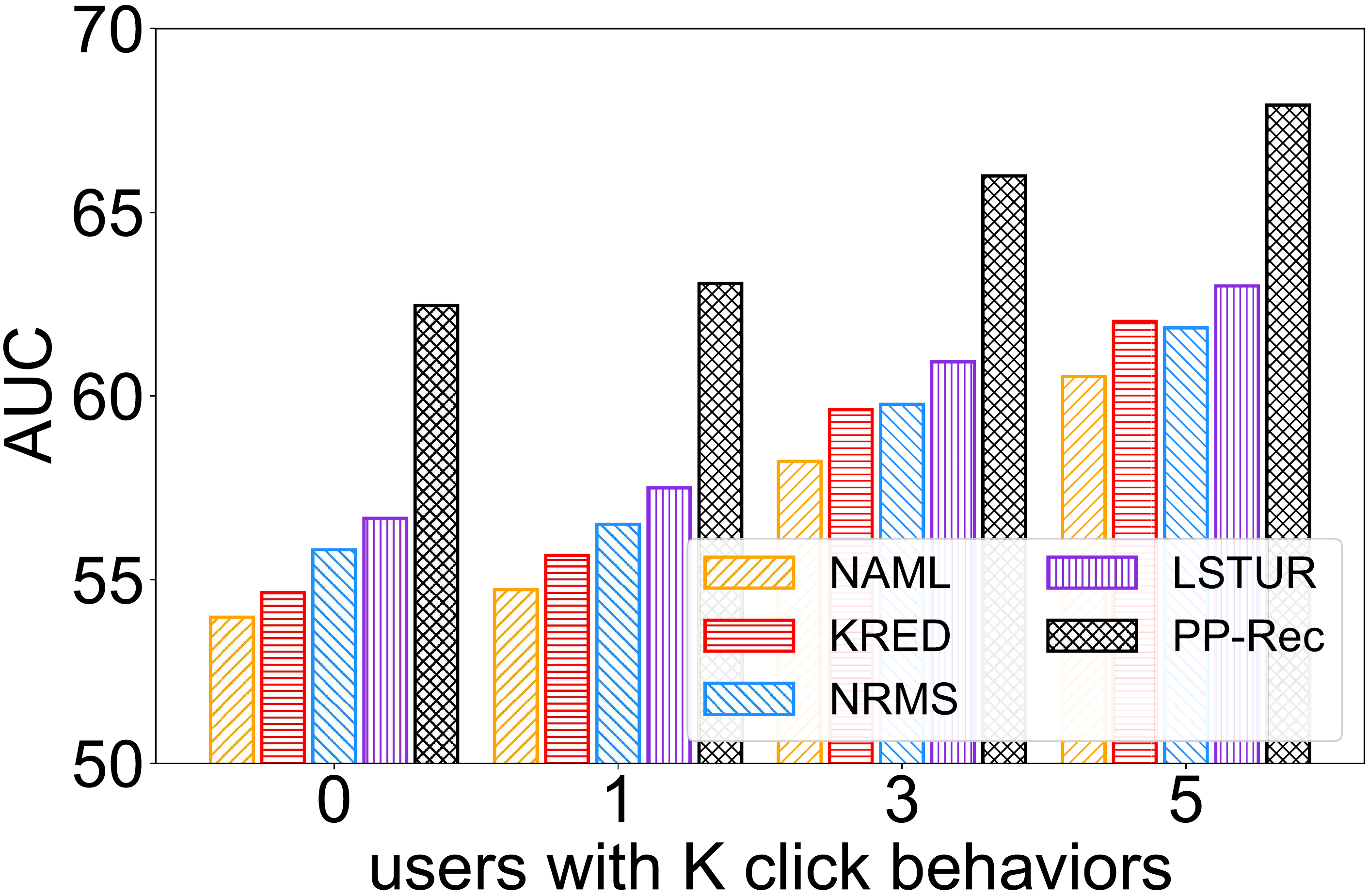}}
    \caption{Performance on cold-start users.}

    \label{fig.cold}

\end{figure}

We evaluate the performance of \textit{PP-Rec} and several personalized methods on news recommendation for cold-start users.
We compare \textit{PP-Rec} with \textit{NAML}, \textit{KRED}, \textit{LSTUR} and \textit{NMRS} since they achieve good performance in Table~\ref{Table.PE}.
We evaluate their performance on recommending news to users with $K\in \{k | k=0,1,3,5\}$ historical clicked news.
In the following sections, we only show experimental results on the \textit{MSN} dataset since results on \textit{MSN} dataset and \textit{Feeds} dataset are similar.
As shown in Fig.~\ref{fig.cold}, \textit{PP-Rec} significantly outperforms other personalized methods.
This is because these personalized methods usually recommend news based on the matching between news and user interests.
However, it is difficult for these methods to accurately model personal interests of cold-start users from their scarce clicks and accurately help them find their interested news.
Different from these methods, \textit{PP-Rec} recommends news based on both personalized interest matching and news popularity.
Popular news usually contains important information and can attract many users with different interests.
Thus, incorporating news popularity into news recommendation can effectively improve the reading experiences of cold-start users.

\subsection{Recommendation Diversity}

	   

\begin{figure}
    \centering
    \resizebox{0.45\textwidth}{!}{
    \includegraphics{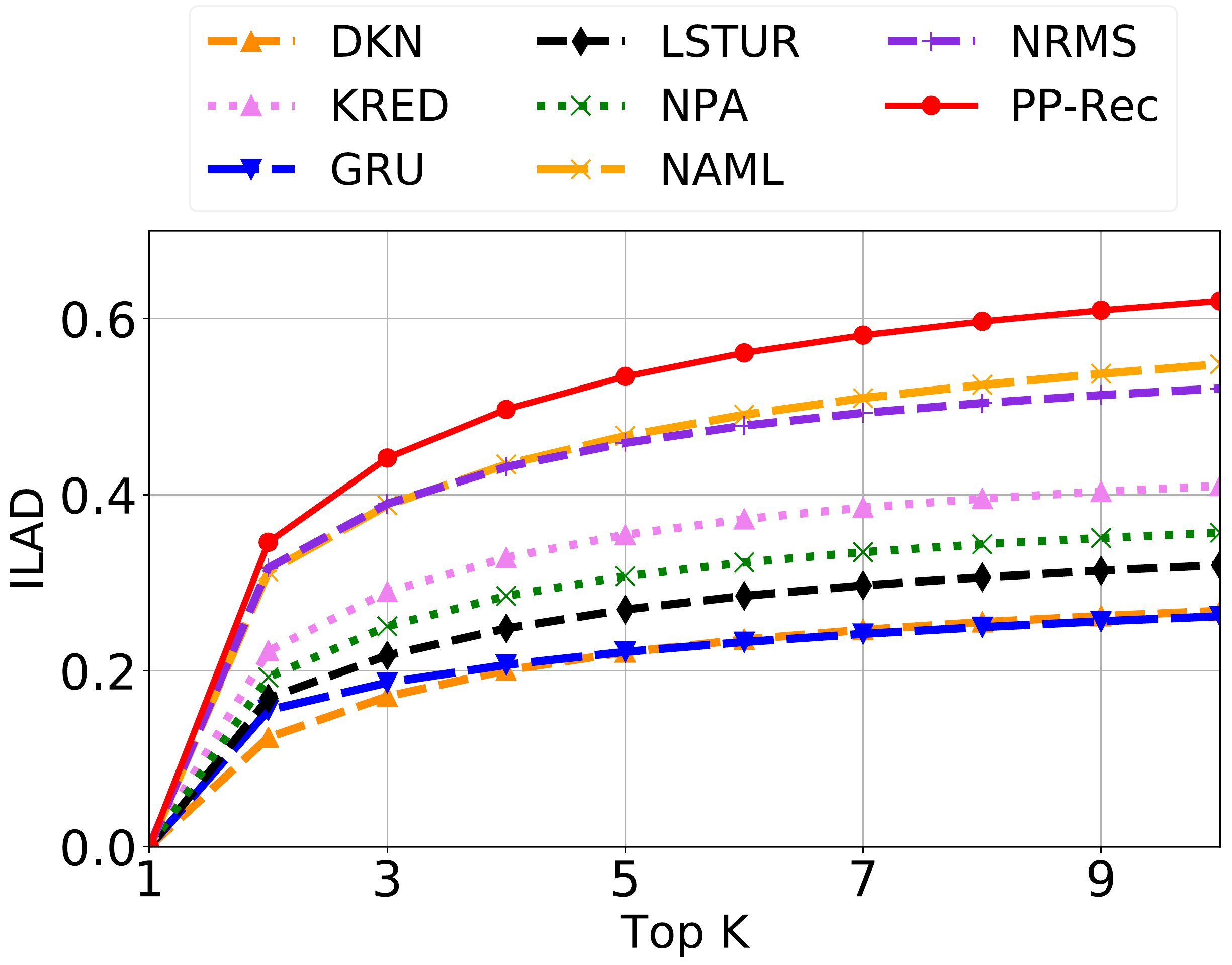}
    }
   \caption{Intra-list average distance of news recommended by different methods.}
    \label{fig.diversity.ilad}
\end{figure}
\begin{figure}
    \centering
    \resizebox{0.45\textwidth}{!}{
    \includegraphics{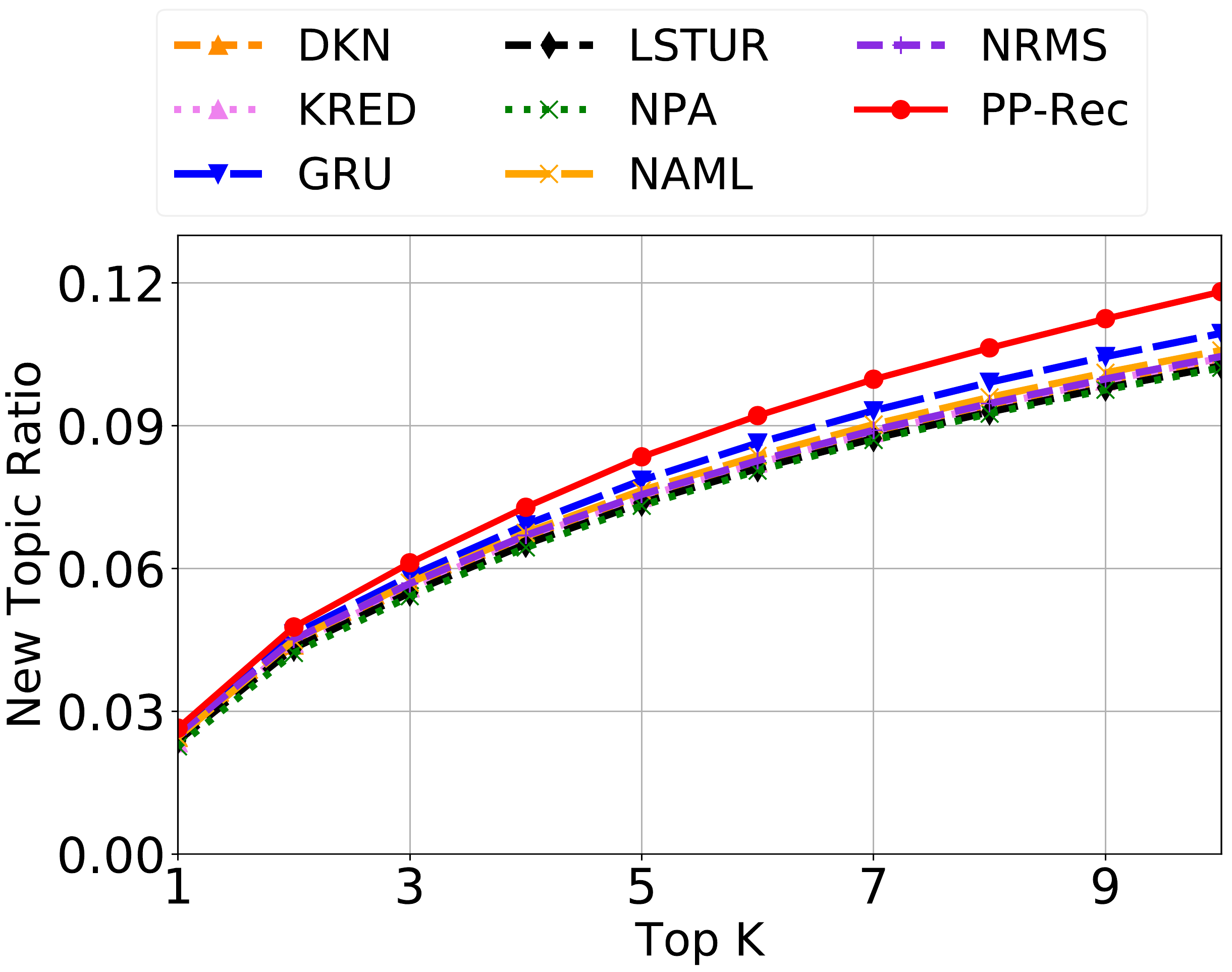}
    }
   \caption{New topic ratio of news recommended by different methods.}
    \label{fig.diversity.topic}
\end{figure}

In this section, we evaluate the recommendation diversity of \textit{PP-Rec} and other personalized methods.
We use two metrics, i.e., intra-list average distance and new topic ratio, to measure the diversity of the top $K$ ($K\in \{k|k=1,...,10\}$) recommended news.
The former is used to measure the average distance between recommended news based on their representations, which is widely used in previous works~\cite{zhang2008avoiding,chen2018fast}.
The second one is used to measure the topic similarity between recommended news and users' historical clicked news.
It counts the number of topics of the top $K$ recommended news which are clicked and are not included in topics of users' historical clicked news.
Besides, we use $K$ to normalize the number.
Fig.~\ref{fig.diversity.ilad} and~\ref{fig.diversity.topic} show that \textit{PP-Rec} can consistently improve the recommendation diversity.
This is because these personalized methods recommend news to users based on the matching between news and user interest inferred from clicked news, making the recommended news tend to be similar to users' consumed news.
Different from these methods, \textit{PP-Rec} incorporates news popularity into news recommendation.
Besides the news which is related to user interest, \textit{PP-Rec} can also recommend popular news, which are very diverse in content and topics, to users.
Thus, \textit{PP-Rec} can enhance recommendation diversity.

\subsection{Ablation Study}

\begin{figure}
    \centering
    \resizebox{0.45\textwidth}{!}{
    \includegraphics{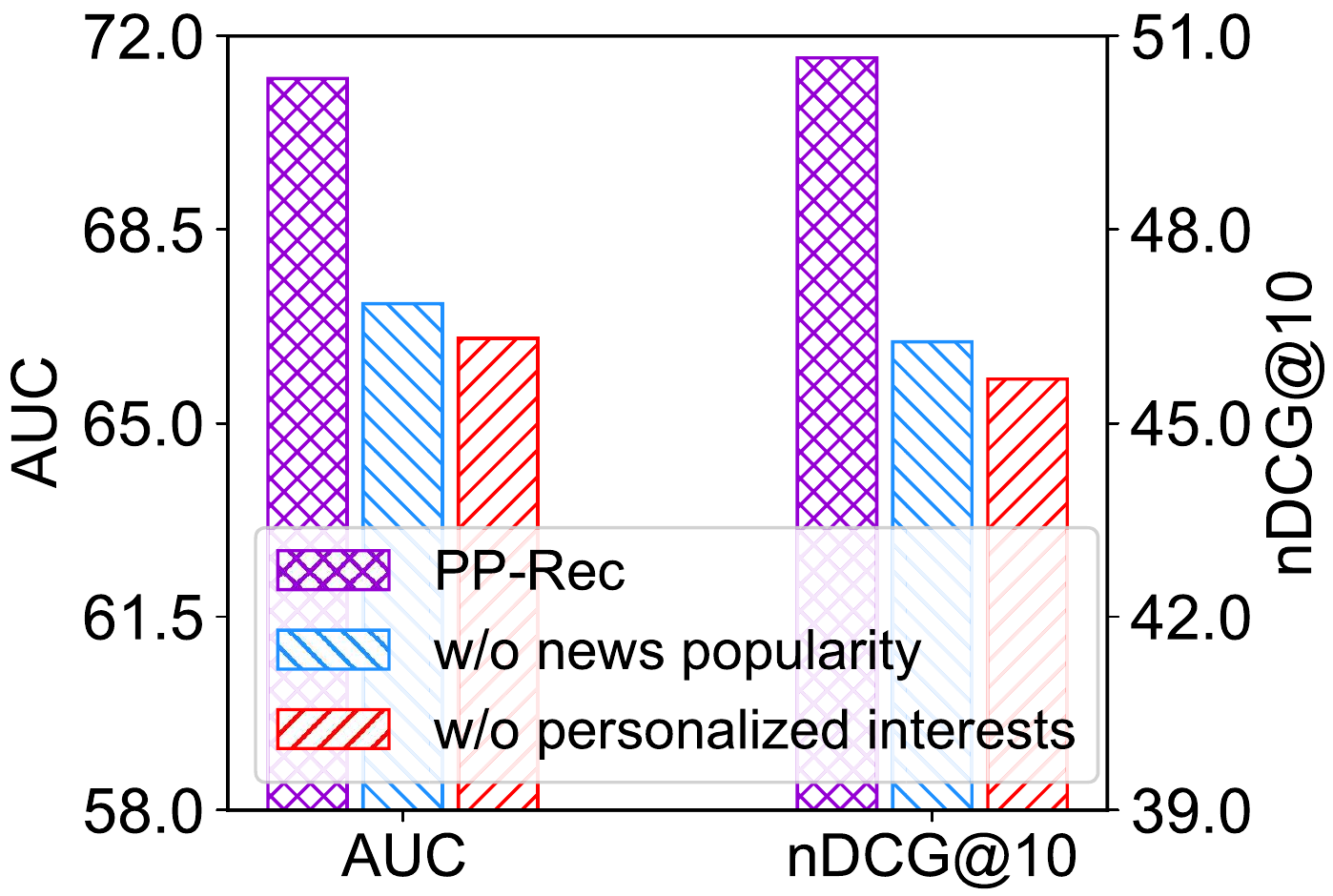}
    }
    \caption{Effectiveness of personalized matching score and news popularity score.}
    \label{fig.score}
\end{figure}

In this section, we conduct several ablation studies on \textit{PP-Rec}.
First, we verify the effectiveness of the two scores for candidate news ranking, i.e., news popularity score and personalized matching score, by removing them individually from \textit{PP-Rec}.
The experimental results are shown in Fig.~\ref{fig.score}.
We have two findings from the results.
First, after removing the news popularity score, the performance of \textit{PP-Rec} declines.
This is because \textit{PP-Rec} incorporates news popularity into news recommendation via this score.
In addition, popular news usually contains important information and can attract many users with different interests.
Thus, recommending popular news can improve news recommendation accuracy.
Second, removing the personalized matching score also hurts the recommendation accuracy.
This is because this score measures user interest in news and incorporates personalized matching into news recommendation in \textit{PP-Rec}.
Since users like to click news related to their personalized interests, recommending users' interested news can effectively improve recommendation accuracy.

\begin{figure}
    \centering
    \resizebox{0.45\textwidth}{!}{
    \includegraphics{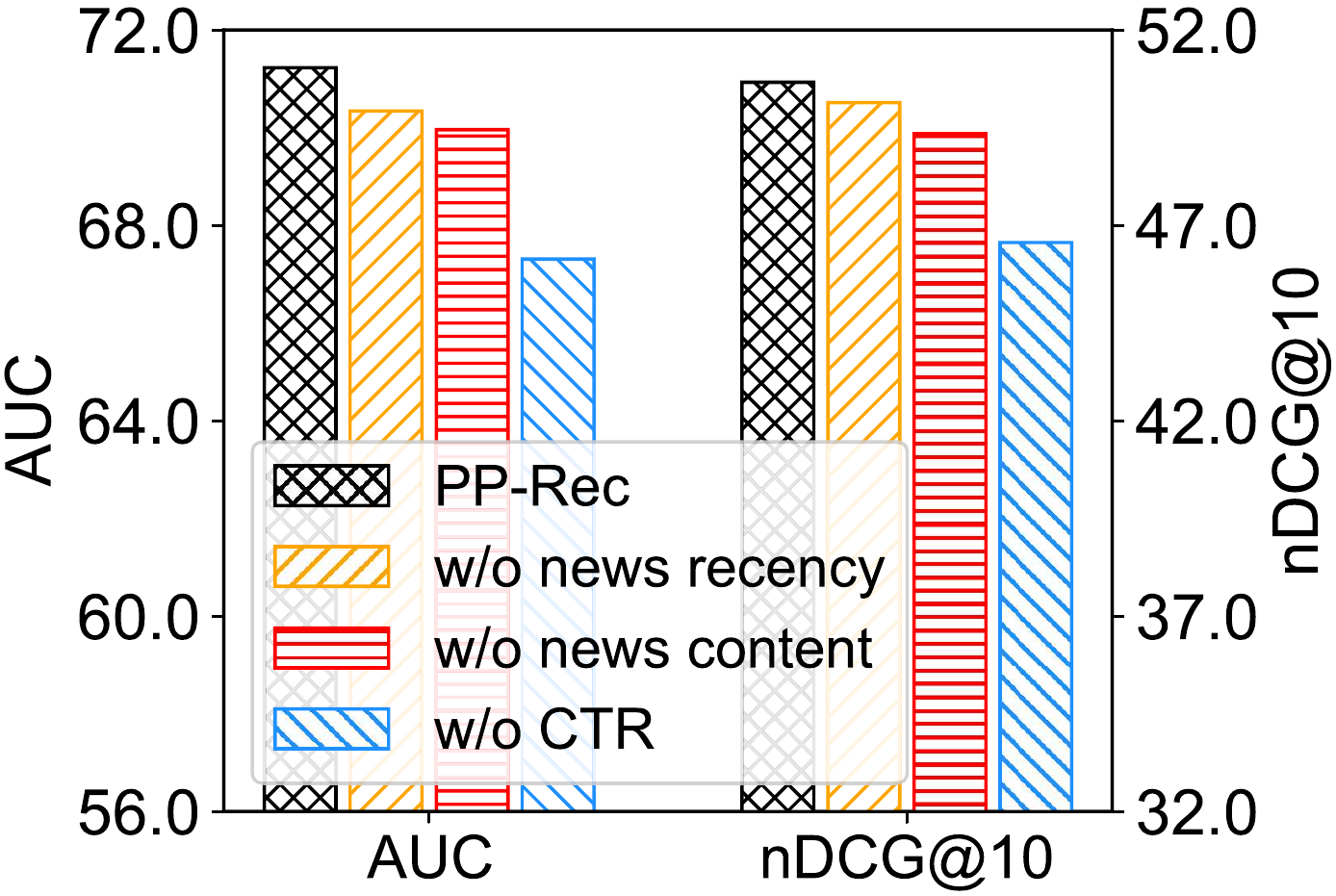}
    }
    \caption{Effectiveness of different information used for news popularity prediction.}

    \label{fig.popularity}
\end{figure}

\begin{figure*}
    \centering
    \resizebox{0.99\textwidth}{!}{
    \includegraphics{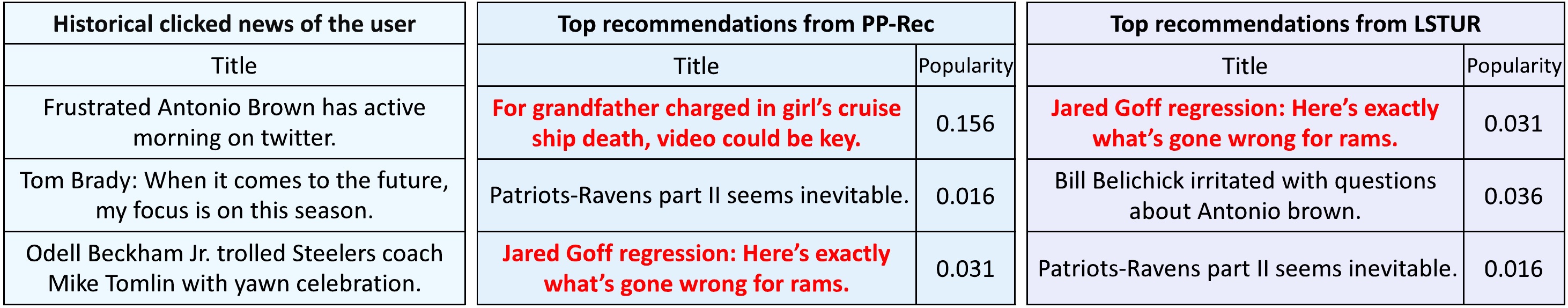}
    }
    \caption{Top news recommended by \textit{PP-Rec} and \textit{LSTUR}. The clicked news are in red and bold.}
    \label{fig.case}
\end{figure*}

Next, as shown in Fig.~\ref{fig.popularity}, we conduct an ablation study to verify the effectiveness of different information in the \textit{time-aware news popularity predictor} by removing them individually.
We have several observations from the results.
First, removing news recency makes the performance of \textit{PP-Rec} decline.
This is because news popularity usually dynamically changes, and popular news will become unpopular once its information is expired.
Since news recency can reflect the freshness of news information, incorporating it makes the news popularity modeling more accurate.
Second, the performance of \textit{PP-Rec} without news content also declines.
This is because after removing it, \textit{PP-Rec} predicts news popularity based on the near real-time CTR and recency.
However, it usually takes some time to accumulate enough impressions to calculate accurate CTR.
Thus, removing the news content makes \textit{PP-Rec} cannot effectively model the popularity of news just published.
Third, \textit{PP-Rec} performs worse without the near real-time CTR.
This is because near real-time CTR effectively measures the click probability of the news based on the behaviors of a large number of users in the recent period. 
Thus, removing the near real-time CTR makes it \textit{PP-Rec} lose much useful information for modeling the dynamic news popularity.

\subsection{Case Study}

We conduct a case study to show the effectiveness of \textit{PP-Rec}.
We compare \textit{PP-Rec} with \textit{LSTUR} since \textit{LSTUR} can achieve the best performance among baseline methods on the \textit{MSN} dataset.
In Fig.~\ref{fig.case}, we list top 3 news recommended by two methods to a randomly sampled user and their normalized popularity predicted by \textit{PP-Rec}.
We also list user's clicked news.
First, we find that the user clicked a news on football, which is recommended by both \textit{LSTUR} and \textit{PP-Rec}.
This is because the user has previously clicked three news on football, which indicates the user is interested in football.
Thus, both \textit{LSTUR} and \textit{PP-Rec} recommend that news based on the personal interest of this user.
Second, the user did not click other news on football recommended by \textit{PP-Rec} and \textit{LSTUR}.
This may be because recommending too much news with similar information may make users feel bored, making the user only click a part of them.
This inspires us that recommending news with diverse information may help improve users' reading experience.
Third, the user clicked a news on crime, which is only recommended by \textit{PP-Rec}.
This is because it is hard to predict user's interests in criminal events from her clicks, making it difficult for \textit{LSTUR} to recommend this news.
Different from \textit{LSTUR}, \textit{PP-Rec} recommends news based on both personal user interest and news popularity.
\textit{PP-Rec} successfully predicts that this news is popular and recommends it.
This case shows that \textit{PP-Rec} can improve the recommendation accuracy and enhance the recommendation diversity by incorporating news popularity.

\section{Conclusion}

In this paper, we propose a new news recommendation method named \textit{PP-Rec} to alleviate the cold-start and diversity problems of personalized news recommendation, which can consider both the personal interest of users and the popularity of candidate news.
In our method, we rank the candidate news based on the combination of a personalized matching score and a news popularity score.
We propose a unified model to predict time-aware news popularity based on news content, recency, and near real-time CTR.
In addition, we propose a knowledge-aware news encoder to generate news content embeddings from news texts and entities, and a popularity-aware user encoder to generate user interest embeddings from the content and popularity of clicked news.
Extensive experiments on two real-world datasets constructed by logs of commercial news websites and feeds in Microsoft validate that our method can effectively improve the accuracy and diversity of news recommendation.

\section*{Acknowledgments}
This work was supported by the National Natural Science Foundation of China under Grant numbers U1936208, U1936216, U1836204, and U1705261.
We are grateful to Xing Xie, Tao Di, and Wei He for their insightful comments and discussions.

\bibliography{anthology}
\bibliographystyle{acl_natbib}


\end{document}